\documentclass[%
 preprint,
 amsmath,amssymb,
 aps,
 pre,
showkeys
]{revtex4-2}

\usepackage{graphicx}
\usepackage{dcolumn}
\usepackage{bm}
\usepackage{hyperref}
\usepackage{xcolor}

\begin{document}

\title{The condensed fraction of a homogeneous dilute Bose gas within the improved Hartree-Fock approximation}

\author{Nguyen Van Thu}
\affiliation{Department of Physics, Hanoi Pedagogical University 2, Hanoi 10000, Vietnam}
\email{nvthu@hpu2.edu.vn}

\author{Jonas Berx}
\affiliation{Institute for Theoretical Physics, KU Leuven, B-3001 Leuven, Belgium}
\email{jonas.berx@kuleuven.be}

\date{\today}

\begin{abstract}
Motivated by the recent experiment [{\it R. Lopes et. al., Phys. Rev. Lett. {\bf119}, 190404 (2017)}] with a homogeneous Bose gas, we investigate a homogeneous dilute Bose gas to calculate the quantum depletion density. By means of the  Cornwall-Jackiw-Tomboulis effective action approach within an improved Hartree-Fock approximation, the condensed fraction is recovered in a simpler manner and compared with corresponding findings in experimental data. Additionally, higher-order terms are taken into account for several physical quantities, in particular for the chemical potential and free energy density.
\end{abstract}


\maketitle


\section{Introduction\label{sec1}}

It is well-known that a number of atoms in a Bose gas will be condensed as the system is cooled to the critical temperature \cite{Bose,Einstein} and results in the formation of a Bose-Einstein condensate (BEC). The literature on this exotic state of matter has been rapidly developing since a BEC was created in experiments \cite{Anderson,Davis}.  In a dilute BEC, essentially all atoms occupy the same quantum state and the condensate can be described in terms of a mean-field theory, which is similar to the Hartree-Fock theory for atoms \cite{Pethick}. Theoretically, all of the atoms will be in the ground state at zero temperature \cite{Pitaevskii}. In this situation, the ground state is described by a wave function, which is the solution of the Gross-Pitaevskii (GP) equation \cite{Gross,Pitaevskii1}.

This absolute zero temperature can however never be reached due to quantum fluctuations in the BEC, where some particles with nonzero momentum reside in excited states instead of the ground state, even at zero absolute temperature \cite{Pethick}. These particles get pushed out of the condensate. This phenomenon is called quantum depletion. The number of atoms in the remaining condensate fraction due to quantum depletion was first studied in 1947 by N. N. Bogoliubov \cite{Bogolyubov} up to order $1/2$ in the gas parameter by using the second quantization formalism. The main idea is based on a quantum description, where the particle operators are transformed into quasi-particle operators, yielding an explicit diagonalization of the quantum Hamiltonian. In 1997, using the Bogoliubov theory and the semiclassical approach, the authors of Ref. \cite{Dalfovo} investigated the quantum depletion in a Bose gas confined by a harmonic trap. In the case of a homogeneous Bose gas their result corresponded exactly with Bogoliubov's result. Recently, this result was reproduced by S. Stringari \cite{Stringari} within the GP theory. However, all of the above methods involve the use of many complicated calculations. The main purpose of the present paper is to provide a simpler method to recover the condensed fraction in a dilute non-relativistic Bose gas with depletion and compare with experimental data.

We set the stage for our calculations by starting with a dilute Bose gas described by the following Lagrangian density \cite{Pethick},
\begin{equation}
{\cal L}=\psi^*\left(-i\hbar\frac{\partial}{\partial t}-\frac{\hbar^2}{2m}\nabla^2\right)\psi-\mu\left|\psi\right|^2+\frac{g}{2}\left|\psi\right|^4,\label{eq:1}
\end{equation}
wherein $\hbar$, $m$ and $\mu$ are respectively the reduced Planck constant, the atomic mass and the chemical potential. The field operator $\psi(\vec{r},t)$ depends on both the coordinate $\vec{r}$ and time $t$. The strength of the interaction between the atoms is determined by the coupling constant $g=4\pi\hbar^2a_s/m$, which is expressed in terms of the $s$-wave scattering length $a_s$ by making use of the Born approximation. Now, thermodynamic stability requires that $g>0$, i.e., the boson interactions are repulsive.

This paper is organised as follows. In Section \ref{sec:2} we calculate the gap and Schwinger-Dyson (SD) equations for a single Bose gas in the improved Hartree-Fock (IHF) approximation by first recapitulating the regular HF method and then calculating these expressions for the Cornwall-Jackiw-Tomboulis effective potential with symmetry-restoring terms, hence the name \textsl{improved} Hartree-Fock method. The condensed fraction is investigated in Section \ref{sec:3} and the results are compared with both related approaches and experimental data. Explicit expressions for the chemical potential, the pressure and the free-energy density are calculated. Finally, we present the conclusions and a future outlook in Section \ref{sec:4}.

\section{The equations of state in the IHF approximation}\label{sec:2}

In this Section, we will establish the equations of state for a Bose gas, which are the gap and Schwinger-Dyson (SD) equations. To do so, we first derive the Cornwall-Jackiw-Tomboulis (CJT) effective potential in the IHF approximation. As was previously mentioned, the condensed fraction is investigated at zero temperature in spite of the fact that the CJT effective action approach is constructed for a finite temperature $T>0$. The temperature is set to be zero at the end of the calculations.

Let $\psi_0$ be the expectation value of the field operator in the tree-approximation, the GP potential is then taken from \eqref{eq:1}
\begin{equation}
V_{GP}=-\mu\psi_0^2+\frac{g}{2}\psi_0^4.\label{eq:VGP}
\end{equation}
Note that henceforth the system is considered without any external fields. Furthermore, the non-macroscopic part of the condensate moves as a whole so that the lowest energy solution $\psi_0$ is real and plays the role of the order parameter. Minimizing the potential \eqref{eq:VGP} with respect to the order parameter, one arrives at the gap equation
\begin{equation}
\psi_0(-\mu+g\psi_0^2)=0,\label{eq:gaptree}
\end{equation}
and hence, for the broken phase
\begin{equation}
\psi_0^2=\frac{\mu}{g}.\label{eq:psi0}
\end{equation}
In order to factor in the HF approximation, the complex field operator $\psi$ should first be decomposed in terms of the order parameter $\psi_0$ and two real fields $\psi_1$ and $\psi_2$, which are associated with quantum fluctuations of the field \cite{Andersen}, i.e.,
\begin{equation}
\psi\rightarrow \psi_0+\frac{1}{\sqrt{2}}(\psi_1+i\psi_2).\label{eq:shift}
\end{equation}
Plugging equation \eqref{eq:shift} into the Lagrangian density \eqref{eq:1}, the interaction Lagrangian density in the HF approximation is obtained
\begin{equation}
{\cal L}_{int}=\frac{g}{2}\psi_0\psi_1(\psi_1^2+\psi_2^2)+\frac{g}{8}(\psi_1^2+\psi_2^2)^2.\label{eq:Lint}
\end{equation}
In the tree approximation one has the gap equation \eqref{eq:psi0} and the inverse propagator or Green's function
\begin{equation}
D_0^{-1}(k)=\left(
              \begin{array}{cc}
                \frac{\hbar^2k^2}{2m}-\mu+3g\psi_0^2 & -\omega_n \\
                \omega_n &  \frac{\hbar^2k^2}{2m}-\mu+g\psi_0^2\\
              \end{array}
            \right),\label{protree}
\end{equation}
with $\vec{k}$ being the wave vector. The $n$th Matsubara frequency for bosons is defined as $\omega_n=2\pi n/\beta$ where $\beta=1/k_BT$ and $n\in{\mathbb{Z}}$ with $k_B$ being the Boltzmann constant. By combining the gap equation \eqref{eq:gaptree} and the inverse propagator (\ref{protree}), the latter reduces to
\begin{equation}
D_0^{-1}(k)=\left(
              \begin{array}{cc}
                \frac{\hbar^2k^2}{2m}+2g\psi_0^2 & -\omega_n \\
                \omega_n &  \frac{\hbar^2k^2}{2m}\\
              \end{array}
            \right).\label{eq:protree1}
\end{equation}
The Bogoliubov dispersion relation can be obtained by requiring that the determinant of the inverse propagator \eqref{eq:protree1} vanishes, i.e.,  $\det D_0^{-1}(k)=0$ \cite{Floerchinger}. The result is
\begin{equation}
E^{\text{(tree)}}(k)=\sqrt{\frac{\hbar^2k^2}{2m}\left(\frac{\hbar^2k^2}{2m}+2g\psi_0^2\right)}.\label{dispertree}
\end{equation}
For small wave vectors $\vec{k}$, this equation is gapless and linear and indicates the spontaneous $U(1)$ symmetry breaking. Due to this symmetry breaking, Nambu-Goldstone bosons (pions) are created. To continue our discussion, we introduce the CJT effective potential in the HF approximation that can be read off from the interaction Lagrangian density \eqref{eq:Lint} in the manner that was pointed out in \cite{Thu,Phat},
\begin{equation}
\begin{split}
V_\beta^{\text{(CJT)}} &=-\mu\psi_0^2+\frac{1}{2}\int_\beta \mbox{tr}\left[\ln G^{-1}(k)+D_0^{-1}(k)G(k)-{1\!\!1}\right]\\
&+\frac{3g}{8}(P_{11}^2+P_{22}^2)+\frac{g}{4}P_{11}P_{22} +\frac{g}{2}\psi_0^4\, ,\label{eq:VHF}
\end{split}
\end{equation}
for which the functions $P_{11}$ and $P_{22}$ are
\begin{subequations}
    \begin{equation}
        \label{eq:P11}
        P_{11}=\int_\beta G_{11}(k)
    \end{equation}
    \begin{equation}
        \label{eq:P22}
        P_{22}=\int_\beta G_{22}(k)
    \end{equation}
\end{subequations}
The Matsubara integrals in these expressions are defined as follows
\begin{equation}
    \label{eq:fk}
    \int_\beta f(k)=\frac{1}{\beta}\sum_{n=-\infty}^{+\infty}\int\frac{d^3\vec{k}}{(2\pi)^3}f(\omega_n,\vec{k})\, .
\end{equation}
Here $G(k)$ is the propagator or Green's function in the HF approximation, which can be obtained by minimizing the CJT effective potential \eqref{eq:VHF} with respect to the elements of the propagator. Performing these calculations results in the following expression for the inverse propagator
\begin{equation}
G^{-1}(k)=D_0^{-1}(k)+\Pi,\label{eq:r11}
\end{equation}
in which
\begin{equation}
\Pi=\left(
              \begin{array}{cc}
                \Pi_1& 0 \\
                0 & \Pi_2\\
              \end{array}
            \right),\label{eq:r12}
\end{equation}
with the matrix entries $\Pi_1$ and $\Pi_1$ the self-energies that can be constructed from \eqref{eq:P11} and \eqref{eq:P22}, i.e.,
\begin{subequations}
    \begin{equation}
        \label{eq:r13}
        \Pi_1=\frac{3g}{2}P_{11}+\frac{g}{2}P_{22}
    \end{equation}
    \begin{equation}
        \label{eq:r14}
        \Pi_2=\frac{g}{2}P_{11}+\frac{3g}{2}P_{22}\, .
    \end{equation}
\end{subequations}
The gap equation in the HF approximation can now be found by minimizing the CJT effective potential \eqref{eq:VHF} with respect to the order parameter $\psi_0$, i.e.,
\begin{equation}
-\mu+g\psi_0^2+\Pi_1=0.\label{eq:r15}
\end{equation}
Combining equations \eqref{eq:r11}-\eqref{eq:r15}, one has the inverse propagator in the HF approximation
\begin{equation}
\label{eq:inverse_G_CJT}
G^{-1}(k)=\left(
              \begin{array}{cc}
                \frac{\hbar^2k^2}{2m}-\mu+3g\psi_0^2+\Pi_1 & -\omega_n \\
                \omega_n &  \frac{\hbar^2k^2}{2m}-\mu+g\psi_0^2+\Pi_2\\
              \end{array}
            \right),
\end{equation}
and consequently the dispersion relation in this approximation is
\begin{widetext}
    \begin{equation}
        E^{\text{(HF)}}(k)=\sqrt{\left(\frac{\hbar^2k^2}{2m}-\mu+3g\psi_0^2+\Pi_1\right)\left(\frac{\hbar^2k^2}{2m}-\mu+g\psi_0^2+\Pi_2\right)}.\label{eq:r16}
    \end{equation}
\end{widetext}

Equations \eqref{eq:r15} and \eqref{eq:r16} show that the Goldstone theorem fails in this HF approximation.

To restore the Goldstone boson, we now employ the method developed in \cite{Ivanov}. In this way, a phenomenological symmetry-restoring extra term
\begin{equation}
\Delta V=-\frac{g}{4}(P_{11}^2+P_{22}^2)+\frac{g}{8}P_{11}P_{22},\label{extra}
\end{equation}
will be added to the CJT effective potential \eqref{eq:VHF}. Let the inverse propagator in the IHF approximation be denoted as $D^{-1}_{\text{(IHF)}}(k)$, the CJT effective potential \eqref{eq:VHF} now becomes
\begin{eqnarray}
    \label{eq:VIHF}
        \widetilde{V}_\beta^{\text{(CJT)}}&&=\frac{1}{2}\int_\beta \mbox{tr}\left[\ln D^{-1}_{\text{(IHF)}}(k)+D_0^{-1}(k)D_{\text{(IHF)}}(k)-{1\!\!1}\right]\nonumber\\
&&+\frac{g}{8}(P_{11}^2+P_{22}^2)+\frac{3g}{8}P_{11}P_{22} -\mu\psi_0^2 +\frac{g}{2}\psi_0^4.
\end{eqnarray}
Similarly, by repeating step by step all of the calculations from the previous discussion for the CJT effective potential in the IHF approximation \eqref{eq:VIHF}, one arrives at the gap equation
\begin{equation}
-\mu+g\psi_0^2+\Sigma_1=0,\label{eq:gap}
\end{equation}
and the SD equation
\begin{equation}
M^2=-\mu+3g\psi_0^2+\Sigma_2,\label{eq:SD}
\end{equation}
respectively. In these equations, the self-energies $\Sigma_1$ and $\Sigma_2$ are denoted by
\begin{subequations}
    \begin{equation}
        \label{eq:sig1}
        \Sigma_1=\frac{3g}{2}P_{11}+\frac{g}{2}P_{22}
    \end{equation}
    \begin{equation}
        \label{eq:sig2}
        \Sigma_2=\frac{g}{2}P_{11}+\frac{3g}{2}P_{22}\, .
    \end{equation}
\end{subequations}

Combining equations \eqref{eq:VIHF}-\eqref{eq:sig2}, one can once again calculate the inverse propagator, now in the IHF approximation,i.e.,
\begin{equation}
D^{-1}(k)=\left(
              \begin{array}{lr}
                \frac{\hbar^2k^2}{2m}+M^2 & -\omega_n \\
                \omega_n & \frac{\hbar^2k^2}{2m} \\
              \end{array}
            \right).\label{eq:proIHF}
\end{equation}
Hence, the resulting dispersion relation is
\begin{equation}
E^{\text{(IHF)}}(k)=\sqrt{\frac{\hbar^2k^2}{2m}\left(\frac{\hbar^2k^2}{2m}+M^2\right)}.\label{disperIHF}
\end{equation}
Clearly, the Goldstone boson is restored in this approximation. This is precisely the reason why this approximation is called the \textsl{improved} Hartree-Fock approximation.

The momentum integrals in the IHF approximation are obtained from equations \eqref{eq:P11} and \eqref{eq:P22} after replacing $G(k)$ by $D(k)$. Using the following formula \cite{Schmitt},
\begin{equation}
\sum_{n=-\infty}^{+\infty}\frac{1}{\omega_n^2+E^2(k)}=\frac{\beta}{2E(k)}\left[1+\frac{2}{e^{\beta E(k)}-1}\right],
\end{equation}
and afterwards letting the temperature tend to zero, it is easy to check that in this limit these momentum integrals have the form
\begin{subequations}
    \begin{equation}
        \label{eq:tichphan1}
        P_{11}=\frac{1}{2}\int\frac{d^3\vec{k}}{(2\pi)^3}\sqrt{\frac{\hbar^2k^2/2m}{\hbar^2k^2/2m+M^2}}
    \end{equation}
    \begin{equation}
        \label{eq:tichphan2}
        P_{22}=\frac{1}{2}\int\frac{d^3\vec{k}}{(2\pi)^3}\sqrt{\frac{\hbar^2k^2/2m+M^2}{\hbar^2k^2/2m}}\, .
    \end{equation}
\end{subequations}

The gap and SD equations \eqref{eq:gap} and \eqref{eq:SD}, together with the momentum integrals \eqref{eq:tichphan1} and \eqref{eq:tichphan2} form the equations of state, which govern the variation of all quantities of the system.

We will now look for the quantum fluctuations, and therefore calculate the condensed fraction of the dilute Bose gas.

\section{The condensed fraction of a homogeneous dilute Bose gas\label{sec:3}}

Let us first investigate the quantum depletion density in the IHF approximation. Note that the pressure is defined as the negative of the CJT effective potential \eqref{eq:VIHF} at the minimum, i.e. satisfying both the gap and SD equations
\begin{equation}
{\cal P}=-\widetilde{V}_\beta\bigg|_{\mbox{minimum}}\equiv -\widetilde{{\cal V}}_\beta^{\text{(CJT)}}.\label{eq:press}
\end{equation}
Now, substituting equations \eqref{eq:gap} and \eqref{eq:SD} into \eqref{eq:VIHF}, one has
\begin{eqnarray}
    \label{eq:V1}
        \widetilde{{\cal V}}_\beta^{\text{(CJT)}}&&=-\mu\psi_0^2+\frac{g}{2}\psi_0^4+\frac{1}{2}\int_\beta \mbox{tr}\ln D^{-1}_{\text{(IHF)}}(k)\\
        &&+\frac{1}{2}(3g\psi_0^2-\mu-M^2)P_{11}\nonumber
    +\frac{1}{2}(g\psi_0^2-\mu)P_{22}\\
    &&+\frac{g}{8}(P_{11}^2+P_{22}^2)+\frac{3g}{8}P_{11}P_{22}.
\end{eqnarray}

One obtains the condensate density in the IHF approximation from differentiating the pressure with respect to the chemical potential, i.e.,
\begin{equation}
n=\frac{\partial {\cal P}}{\partial \mu}.\label{eq:rho}
\end{equation}
Combining equations \eqref{eq:press}-\eqref{eq:rho}, the condensate density is expressed in terms of the order parameter $\psi_0$ and the momentum integrals $P_{11}$ and $P_{22}$ \cite{Phat},
\begin{equation}
n=\psi_0^2+\frac{1}{2}(P_{11}+P_{22}).\label{eq:n}
\end{equation}
Similarly, the chemical potential can be expressed in terms of the bulk condensate density $n_0$
\begin{equation}
\mu=gn_0+gP_{11}.\label{eq:chemical}
\end{equation}
Physically, the first term in equation \eqref{eq:chemical} is the chemical potential in the mean field theory while the second one originates from the double-bubble diagrams in the IHF approximation.

\begin{figure}[!ht]
    \centering
    \includegraphics[width = 0.95\linewidth]{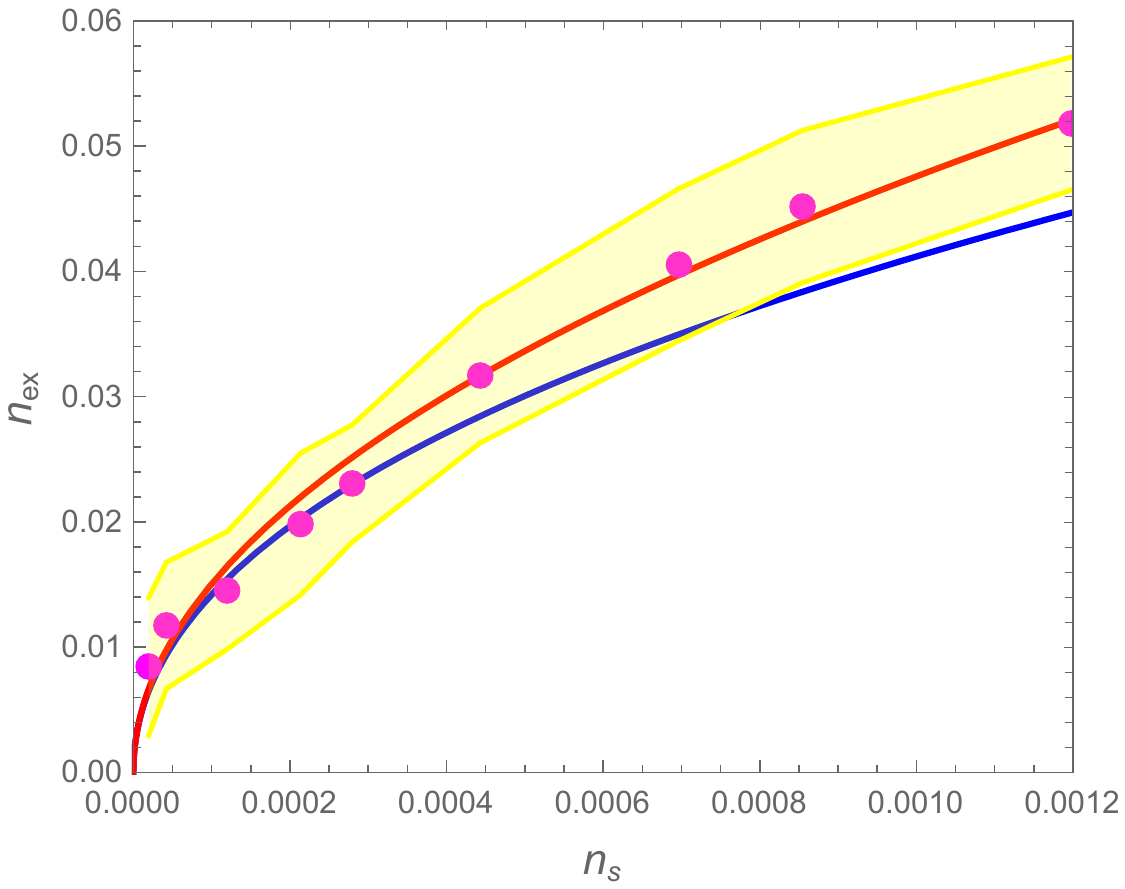}
    \caption{(Color online) The condensed fraction $n_{ex}/n_0$ of a homogeneous Bose gas as a function of the gas parameter $n_s$. The red line corresponds to Bogoliubov's result and to the first term on the right-hand side of equation \eqref{eq:ratio}. The blue line is the numerical solution for the condensed fraction. The magenta dots are experimental data \cite{Lopes} together with a yellow error band.}\label{fig:f1}
\end{figure}

The density of condensate depletion is defined as the number of particles in excited states per unit volume \cite{Pethick}. Based on equation \eqref{eq:n} one can easily see that the condensed fraction for a homogeneous dilute Bose gas is
\begin{equation}
n_{ex}=\frac{1}{2}(P_{11}+P_{22}).\label{eq:ne}
\end{equation}
In order to simplify notation, we first introduce the following dimensionless quantities
\begin{subequations}
    \begin{equation}
        \label{eq:healing_length}
        \xi=\hbar^2/\sqrt{2mgn_0}
    \end{equation}
    \begin{equation}
        \label{eq:reduced_order_parameter}
        \phi_0=\psi_0/\sqrt{n_0}
    \end{equation}
     \begin{equation}
        \label{eq:wave_vector}
        \kappa=k\xi
    \end{equation}
    \begin{equation}
        \label{eq:effective_mass}
        {\cal M}=M/\sqrt{gn_0}\, ,
    \end{equation}
\end{subequations}
with $\xi$ the coherent healing length, $\phi_0$ the reduced order parameter, $\kappa$ the wave vector and ${\cal M}$ the effective mass. The momentum integrals \eqref{eq:tichphan1} and \eqref{eq:tichphan2} can be conveniently rewritten as
\begin{subequations}
    \begin{equation}
        \label{eq:tichphan1a}
        P_{11}=\frac{1}{2\xi^3}\int\frac{d^3\vec{\kappa}}{(2\pi)^3}\frac{\kappa}{\sqrt{\kappa^2+{\cal M}^2}}
    \end{equation}
    \begin{equation}
        \label{eq:tichphan2a}
        P_{22}=\frac{1}{2\xi^3}\int\frac{d^3\vec{\kappa}}{(2\pi)^3}\frac{\sqrt{\kappa^2+{\cal M}^2}}{\kappa}\, .
    \end{equation}
\end{subequations}
The integrations over the dimensionless wave vector in equations \eqref{eq:tichphan1a} and \eqref{eq:tichphan2a} are ultraviolet divergent. With dimensional regularization, these divergences can be avoided \cite{Andersen} and the integrals can be solved. The integral $I_{m,n}$ is
\begin{widetext}
    \begin{equation}
        I_{m,n}({\cal M})=\int\frac{d^d\kappa}{(2\pi)^d}\frac{\kappa^{2m-n}}{(\kappa^2+{\cal M}^2)^{n/2}}\nonumber
        =\frac{\Omega_d}{(2\pi)^d}\Lambda^{2\epsilon}{\cal M}^{d+2(m-n)}\frac{\Gamma\left(\frac{d-n}{2}+m\right)\Gamma\left(n-m-\frac{d}{2}\right)}{2\Gamma\left(\frac{n}{2}\right)},\label{eq:tp}
    \end{equation}
\end{widetext}
where $\Gamma(x)$ is the gamma function, $\Omega_d=2\pi^{d/2}/\Gamma(d/2)$ is the surface area of a $d-$dimensional sphere and $\Lambda$ is a renormalization scale that ensures the integral has the correct canonical dimension. This number is usually absorbed into the measure, hence it will not appear in the results.  Applying \eqref{eq:tp} to \eqref{eq:tichphan1a} and \eqref{eq:tichphan2a} with $d=3$, one finds
\begin{subequations}
    \begin{equation}
        \label{eq:tp11}
        P_{11}=\frac{{\cal M}^3}{6\pi^2\xi^3}
    \end{equation}
    \begin{equation}
        \label{eq:tp22}
        P_{22}=-\frac{{\cal M}^3}{12\pi^2\xi^3}\, .
    \end{equation}
\end{subequations}
To proceed further, we define the gas parameter $n_s=n_0a_s^3$. Inserting equations \eqref{eq:tp11}, \eqref{eq:tp22} and \eqref{eq:chemical} into \eqref{eq:gap} and \eqref{eq:SD}, one arrives at the gap equation
\begin{equation}
-1+\phi_0^2+\frac{2\sqrt{2}n_s^{1/2}}{3\sqrt{\pi}}{\cal M}^3=0,\label{eq:gap1}
\end{equation}
and the SD equation
\begin{equation}
{\cal M}^2=-1+3\phi_0^2-\frac{10\sqrt{2}n_s^{1/2}}{3\sqrt{\pi}}{\cal M}^3,\label{eq:SD1}
\end{equation}
Note that for a dilute Bose gas, the gas parameter $n_s$ has to satisfy the condition $n_s\ll1$ \cite{Andersen}.
Solving equations \eqref{eq:gap1} and \eqref{eq:SD1} and then expanding, one easily finds the dimensionless effective mass
\begin{equation}
{\cal M}=\sqrt{2}-\frac{16\sqrt{2}}{3\sqrt{\pi}}n_s^{1/2}+{\cal O}(n_s)\label{eq:Mm}
\end{equation}
and the reduced order parameter
\begin{equation}
\phi_0^2=1-\frac{8}{3\sqrt{\pi}}n_s^{1/2}+{\cal O}(n_s^2).\label{eq:order}
\end{equation}
Finally, substituting \eqref{eq:Mm} into \eqref{eq:ne} one finds the condensed fraction
\begin{equation}
\frac{n_{ex}}{n_0}=\gamma n_s^{1/2}-\frac{128n_s}{3 \pi }+\frac{2048 n_s^{3/2}}{9 \pi ^{3/2}}+{\cal O}(n_s^{2}),\label{eq:ratio}
\end{equation}
in which $\gamma=\frac{8}{3\sqrt{\pi}}\approx 1.50$. Note that here, in accordance with Ref. \cite{Bogolyubov}, $n_0$ is assumed to be the density of the condensate, which is the square of the expectation value \eqref{eq:psi0} of the field operator in the tree-approximation. The evolution of the condensed fraction versus the gas parameter is shown in Fig. \ref{fig:f1} \footnote{The experimental data is rescaled from Ref. \cite{Lopes} and was provided to us by Raphael Lopes.}.

Recently, Lopes {\it et. al.} have performed an experiment \cite{Lopes} in order to measure the quantum depletion density of a dilute homogeneous Bose gas of $^{39}_{19}$K atoms as a function of the gas parameter $n_s$. The interaction strength was controlled by a magnetic Feshbach resonance \cite{Inouye} with a particle density of $n_0=3.5\times 10^{11}$ cm$^{-3}$ in the lowest hyperfine state $\left|F=1,m_F=1\right>$.  Within a 15\% statistical error and with 20\% systematic effects, the authors found $\gamma=1.5(2)$. This work confirmed Bogoliubov's result and, naturally, verified our result. However, the higher-order terms can be easily retained in our calculations. As a consequence, our results coincide more closely with the experimental data in the extremely small gas parameter region.

Let us remark that, by plugging \eqref{eq:Mm} and \eqref{eq:order} into \eqref{eq:chemical}, one arrives at the expression for the chemical potential in function of the gas parameter with the inclusion of higher-order terms, i.e.,
\begin{equation}
\mu=gn_0\left[1+\frac{32}{3}\left(\frac{n_s}{\pi}\right)^{1/2}-\frac{512}{3}\frac{n_s}{\pi}+\frac{8192}{9}\left(\frac{n_s}{\pi}\right)^{3/2}+{\cal{O}}(n_s^2)\right].\label{eq:po}
\end{equation}
The two first terms in the bracket on the right-hand side of equation \eqref{eq:po} have been found in Ref. \cite{Andersen}, confirming our result.

Finally, one other quantity of interest is the energy density, which is the Legendre transform of the free-energy density $\mathcal{F(\mu)}$, i.e.,
\begin{equation}
\label{eq:e1}
\begin{split}
    {\cal E}(n)&=\mathcal{F(\mu)} + \mu n\\
    &= \mu n -{\cal P}\, .
\end{split}
\end{equation}
In the last line of \eqref{eq:e1} we used the fact that $\mathcal{F(\mu)} = -{\cal P}$. Substituting  \eqref{eq:V1}, \eqref{eq:n} and \eqref{eq:chemical} into \eqref{eq:press} one can express the pressure in term of the condensate density
\begin{equation}
{\cal P}=\frac{g}{2}n^2+gnP_{11}-\frac{g}{2}P_{11}^2-\frac{1}{2}\int_\beta\mbox{tr}\ln D^{-1}_{\text{(IHF)}}(k).\label{eq:p1}
\end{equation}
Combining equations \eqref{eq:e1} and \eqref{eq:p1} yields the energy density
\begin{equation}
{\cal E}=\frac{g}{2}(n^2+P_{11}^2)+\frac{1}{2}\int_\beta\mbox{tr}\ln D^{-1}_{\text{(IHF)}}(k).\label{eq:e2}
\end{equation}
To proceed further, one has to calculate the last term on the right-hand side of \eqref{eq:e2}. Using the rule for the summation of Matsubara frequencies \cite{Schmitt},
\begin{equation}
\frac{1}{\beta}\sum_{n=-\infty}^{n=+\infty}\ln\left[\omega_n^2+E^2(k)\right]=E(k)+\frac{2}{\beta}\ln\left[1-e^{-\beta E(k)}\right],
\end{equation}
one can evaluate this integral, which we call the grand canonical energy density at zero temperature
\begin{equation}
    \begin{split}
        \Omega&\equiv\frac{1}{2}\int_\beta \mbox{tr}\ln D^{-1}_{\text{(IHF)}}(k)\\
        &=\frac{1}{2}\int\frac{d^3\vec{k}}{(2\pi)^3}\sqrt{\frac{\hbar^2k^2}{2m}\left(\frac{\hbar^2k^2}{2m}+M^2\right)},
    \end{split}
\end{equation}
or in dimensionless form
\begin{equation}
\Omega=\frac{gn_{0}}{2\xi^3}\int\frac{d^3\kappa}{(2\pi)^3}\sqrt{\kappa^2(\kappa^2+{\cal M}^2)}.\label{eq:energy}
\end{equation}
Using equation \eqref{eq:tp}, the integral can be performed explicitly. Hence, the grand canonical energy density is
\begin{equation}
\Omega=\frac{gn_0}{30\pi^2\xi^3}{\cal M}^5.\label{eq:energy1}
\end{equation}
From equations \eqref{eq:tp11}, \eqref{eq:Mm}, \eqref{eq:e2} and \eqref{eq:energy1}, the free energy density can be written in term of the gas parameter
\begin{equation}
{\cal E}={\cal P}_0\left[1+\frac{128}{15}\left(\frac{n_s}{\pi}\right)^{1/2}-\frac{1792}{9}\frac{n_s}{\pi}-\frac{20480}{27}\left(\frac{n_s}{\pi}\right)^{3/2}\right],\label{eq:energy2}
\end{equation}
where ${\cal P}_0=gn_0^2/2$ is the pressure in bulk. In the right-hand side of Eq. \eqref{eq:energy2}, the first term is the mean field energy and second one is due to the quantum fluctuations. The result \eqref{eq:energy2} exactly coincides with the corresponding one derived in Ref. \cite{Lee1,Lee2}, which was first obtained by Lee, Huang, and Yang in the late 1950s and was confirmed experimentally by Navon {\it et. al.} \cite{Navon} in 2011.

\section{Conclusion and outlook\label{sec:4}}

In this paper, we have explored the CJT effective action approach to investigate the quantum fluctuations in a homogeneous dilute Bose gas within the framework of the IHF approximation. The known result by Bogoliubov for the condensed fraction has been recovered by means of a simple procedure. A comparison with experimental data shows that our solution coincides almost exactly with the data, especially in the region $n_s < 0.0004$ where the gas parameter is extremely small. Moreover, the CJT effective action approach has been employed to reproduce the relation for the chemical potential and ground state energy of a dilute Bose gas, taking into account the quantum fluctuations. The results coincide exactly with known results that were first calculated in Refs. \cite{Lee1,Lee2}, and can therefore be compared with the experimental result in \cite{Navon}.

It is also very interesting to explore these calculations for Bose-Einstein condensate mixtures. The interspecies interaction is expected to produce some novel results. In addition, thermal fluctuations can be similarly investigated by means of the CJT effective action approach. In addition, this procedure can also be employed to investigate the condensed fraction of a Bose gas confined between two parallel plates and the resulting Casimir effect \cite{Thu}.

\section*{Acknowledgements}

This research is funded by Vietnam National Foundation for Science and Technology Development (NAFOSTED) under grant number 103.01-2018.02. We are very grateful to M. Guilleumas for the useful conversations and to L. Raphael for extensive discussions about the experimental data.

\bibliography{biblio.bib}

\end{document}